# Equilibrium in solid-state electrochemical cells at sample-electrolyte boundary.

Ya.I. Korepanov[1]

[1] D.S. Korzhinsky Institute of Experimental Mineralogy of the Russian Academy of Sciences (IEM RAS)

*e-mail: yakoff@iem.ac.ru*

**Abstract.**

In solid-state electrochemical experiments, the boundary between the sample and the electrolyte plays a crucial role, and the rest of the sample acting as a buffer that maintains a fixed composition. Due to the presence of an electrochemical circuit in the cell (ionic conductivity in the electrolyte and electronic conductivity in the sample system), the chemical potential is equalized at the electrolyte-sample boundary. This leads to an equilibrium state and allows for the measurement of equilibrium values, regardless of the assumptions made by the experimenters.

This article demonstrates the principle of equalizing the electrochemical potential at the electrolyte-sample interface, which explains why achieving equilibrium in the presence of an electrolyte can occur much faster compared to sample synthesis, which often occurs at higher temperatures. This fact can be used to synthesize sample systems or other substances at the required temperatures with a significant reduction in synthesis time.

## 1. Introduction.

Obtaining thermodynamic properties of phases through experimentation is a fundamental task [1-6]. The main criterion for obtaining accurate data is stability and equilibrium in the measurement system. The isobaric potential, also known as the Gibbs free energy, is the most convenient thermodynamic function for describing equilibria in multicomponent systems. This function depends on the independent variables P and T, and it expresses various properties of the system using its derivatives [7,8]. The EMF method [1-4,15-26] is a significant and direct method for studying thermodynamic properties. The technology of the experiment is a difficult and important task for the experimenter. The purpose of this article is to describe the principle of achieving equilibrium in the sample system at the interface with the electrolyte. This article explains the principle and reasons for establishing equilibrium and attempts to show the kinetics of the process.

## 2. Theoretical background.

To construct logical arguments, it is necessary to discuss topics such as equilibrium state, the principle of operation of an EMF cell as the basis for proving the postulated statements.

## 2.1. Equilibrium state.

A logical condition for equilibrium [10-14] is stated as follows: "In any equilibrium heterogeneous system, the chemical potential of each component is the same in all phases."

## 2.2. EMF Cell.

Under the condition of the existence of an ionic conductor with a constant valence and low electronic conductivity, it is possible to create a EMF cell:

$$(-)A|\text{solid state electrolyte A}| A\text{-}B\text{-}C(+) \qquad (1)$$

Here A is the reference system, and A-B-C is the sample system. The obtained EMF values, measured either by a compensation circuit or by a voltmeter with a resistance of the order of $10^{12}$ ohms, correspond with an accuracy of up to a coefficient for the work of transferring one mole of substance A through a solid electrolyte:

$$\mu_A^{reference\ system} - \mu_A^{sample\ system} = -z_A F E$$

Where $z_A$- valence of ions in electrolyte, F- Faradays constant, $E$- measured values of EMF. A fully detailed description of the EMF method can be fined in paper [3] and cited literature.

## 3. Proof.

Consider a hypothetical situation in an electrochemical cell(1).

Let there be 2 points on the surface of the sample with different chemical potentials $\mu_A^1$ и $\mu_A^2$, where $\mu_A^1 > \mu_A^2$, then the system will tend to an equilibrium.

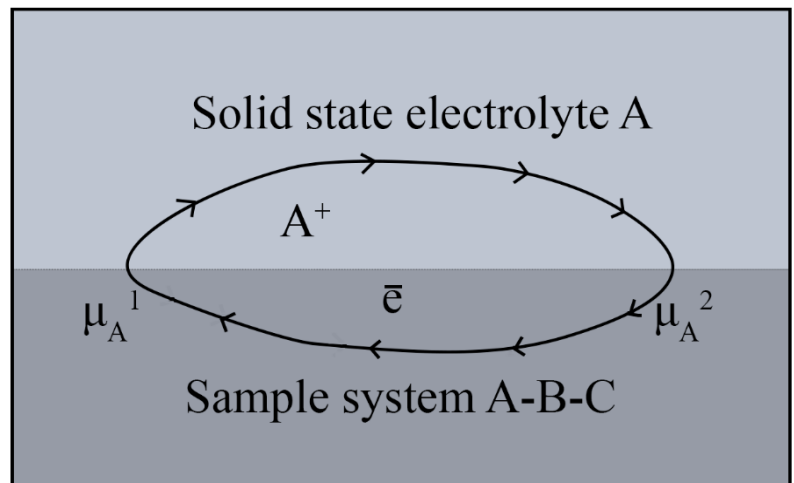

**Fig.1.** Scheme of virtual electrochemical cell A-B-C $\mu_A^1$ |solid state electrolyte $A^+$|A-B-C $\mu_A^2$

The sample system can be quite stable due to the fact that at the temperatures of the experiment, the diffusion kinetics does not allow processes to occur during the experiment (around six months), while the rate of transfer of matter through electrolytes, for example $Ag_4RbI_5$, is significantly higher. The electronic conductivity of the sample system, as a rule, is significant (it is one of the criteria for a correct experiment). As a result, we get a closed electrochemical cell (Fig.1.):

A-B-C $\mu_A^1$ |solid state electrolyte $A^+$|A-B-C $\mu_A^2$

inside the cell (1).

## 4. Discussion.

The fact that non equilibrium state can exist at experimental temperatures for a longer time than the duration of the experiment but the system will tend to an equilibrium. This can be illustrated by comparison the chemical potential to the water level in three interconnected vessels, one of which is linked to the others through a thin capillary tube. In this scenario, the electrolyte serves as a broad tube connecting all the vessels together. As a result, there will be an equilibrium in the system.

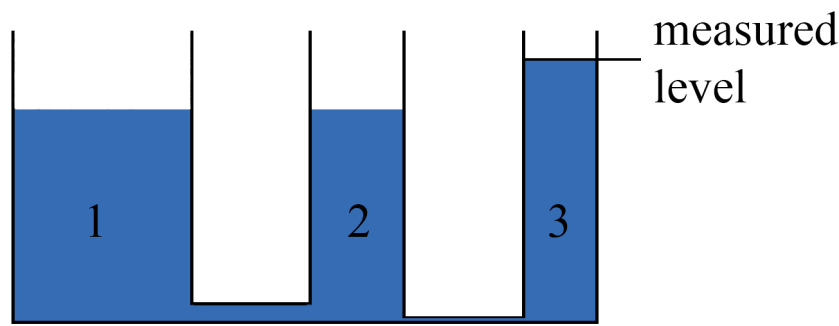

**Fig.2.** Three vessels. Measured level can be higher than in the equilibrium state. Comparison the chemical potential to the water level in three interconnected vessels, one of which is linked to the others through a thin capillary tube.

But as an example [1] – " We assume that the $Ag_2S$-$Au_2S$ cross section is a binary system, and the homogeneity range of uytenbogaardtite and petrovskaite is very narrow (Folmer et al. 1976). All reactions and galvanic cells listed above correspond to the pseudoternary system $Ag_2S$-$Au_2S$-$Ag_xAu_{1-x}$. As the system does not contain pure sulfur, electrum of any fixed composition, as well as pure gold, can exist in Cells A, B, and C. However, to reduce the number of auxiliary data (Au activity in electrum of fixed composition), these cells contained only pure crystalline gold." This citation suggests that the authors assumed that any desired amount of liquid can be added to the third vessel without affecting the measured level (Fig. 2).

Thus, it turns out that for any system in the presence of an electrolyte, the process of substance transfer occurs significantly faster, and therefore equilibrium is reached faster than one would assume. This phenomenon can be used to synthesize substances and achieve an equilibrium state at low temperatures. In such conditions, the synthesis of a substance can take months or years; however, in the presence of an electrolyte, this process can be accomplished in a week or less.

Literature.

r